\documentclass{ws-procs9x6}

\begin{document}

\title{
Effects of FSI, CR and BEC at Small Relative Momenta of Particles 
and W-mass Systematics at LEP}

\author{Nelli Pukhaeva}

\address{ Kansas State University, Manhattan, USA\\
Joint Institute for Nuclear Research, Dubna, Russian Republic}

\author{Amiran Tomaradze}

\address{ Northwestern University, Evanston, Illinois, USA}


\maketitle

\abstracts
{The effects of Final State Interactions(FSI), Colour Reconnection(CR) and
Bose--Einstein Correlations(BEC) at small relative momenta of particles
are discussed. 
The short review of the LEP results on BEC 
at Z-peak, on BEC between particles from different Ws in 
$e^+e^-\rightarrow W^+W^-$ events, and the W-mass systematics due to BEC is given.}

\section{Bose-Einstein Correlations at LEP}


The observation of the enhanced production of like--sign particle pairs
by LEP experiments at Z does not insure that the observed
correlations owe their origin to Bose-Einstein correlations(BEC).
They could, for example, arise
from final state interactions(FSI) between pions with small relative momenta.
(we recommend not to call BEC and/or Colour Reconnection as 'FSI effects' 
as it is generally used).
It turns out that it is possible to discount the possibility of 
visible FSI effects by virtue of the smaller correlations between 
unlike-sign particles, which 
could arise from final state interaction effects too. As was shown at 
LEP, the unlike-sign particle configurations show also 
a correlations.
The like-sign pions necessarily are in an isospin I=2 state and the 
unlike-sign pions are dominantly in the isospin
I=0 state. The scattering length for I=2 is more than a factor three smaller
than the one for I=0. This insures that the FSI
effects for like-sign pion pairs
are expected to be nearly an order of
magnitude smaller than those observed in unlike-sign pions. Even if one
would attribute all the correlations observed for unlike-sign pions to 
final state effects,
then essentially none of the correlation effects observed in like-sign pairs
can therefore be attributed to final state interactions.
As for Colour Reconnection(CR) effects, there is a no evidence 
that any of CR model predicts visible correlations between 
pion pairs at small relative momenta.

On the other hand, it is known that at LEP energies for high
multiplicity events (like $e^+e^-\rightarrow Z \rightarrow hadrons$ and 
$e^+e^-\rightarrow W^+W^-\rightarrow hadrons$),
 as well as a clear effect 
on like--sign particle correlations, the BEC affect unlike--sign particle 
pairs as well.
These so-called 'residual BEC', discussed in details in~\cite{lafferty},
were observed at Z energies by three LEP experiments.(The distortion of
the $\rho^{0}(770)$ line--shape in the mode $\rho^{0}(770)\rightarrow\pi^{+}\pi^{-}$ was observed. These measurements are model independent.)
The influence of BEC to unlike--sign particles 
is also predicted by LUBOEI~\cite{PYTHIA} and by a
'reweighting' BEC model~\cite{lafferty}. 
From the generalised Bose statistics and isospin invariance follows also 
that the pairs with different charges which belong to the same 
isospin multiplet may show BEC 
enhancement.
Thus, one may conclude that the observed at LEP correlations 
for like-sign as well for unlike-sign particle
pairs can be ascribed to Bose--Einstein Correlations.

As a consequence, we recommend do not divide(or subtract) the
plots of like--sign and unlike--sign particle configurations 
in analysis of the Bose-Einstein Correlations effects in 
$e^+e^-\rightarrow Z$ and $e^+e^-\rightarrow W^+W^-$ events.

\section{LEP Results on Correlations Between Particles from different
Ws in  $e^+e^-\rightarrow W^+W^-$ Events}

\subsection{The DELPHI results} 

The ratio $D(Q)$ for DELPHI data obtained using practically
all available statistics,
for model with full and inside Ws BEC using LUBOEI BE$_{32}$~\cite{PYTHIA}, 
are shown in Fig.~1 for like--sign
and for unlike--sign pairs~\cite{newnote}.
The $D$ is the ratio of plots of WW fully 
hadronic to mixed semileptonic events. The $Q$ is defined as
$Q^{2}$=$M_{\pi\pi}^{2}$--4$m_{\pi}^{2}$.
The alternative independent analysis~\cite{newnote} of the same DELPHI data
using a different event selections and mixing method yielded practically 
identical measurements of the $D(Q)$. 

DELPHI reports an effect of inter--W correlations at the level 
of 2.8$\sigma$ for like--sign particle pairs
$\Lambda$(DELPHI)$ =0.142 \pm 0.049 (stat) \pm 0.015 (syst).$
This value of $\Lambda$(DELPHI) has been obtained by fitting
the $D(Q)$ plot  in range of $Q$=0--4.0 GeV using the equation
$D(Q) = N \left( 1 + \delta Q \right) \left( 1 + \Lambda e^{-R Q} \right)$.
 The parameters $N$, $\delta$ and $\Lambda$ were fit parameters,
while the parameter $R$ were fixed to value $R$=0.82 fm, as was obtained
from the model with full BEC tuned to the data at Z-peak.
The value $\Lambda$ for the LUBOEI  BE$_{32}$ with full BEC was 
$\Lambda(full~BEC)$=0.241$\pm$0.009.

DELPHI reports also an effect of inter--W correlations at the level 
of 2.0$\sigma$ for unlike--sign particle pairs.
The value $\Lambda$(DELPHI) for unlike--sign particle pairs was in agreement
with prediction of the  LUBOEI  BE$_{32}$ with full BEC.

\vspace*{-1.5cm}

\begin{figure}[ht]
\centerline{\epsfxsize=3.5in\epsfbox{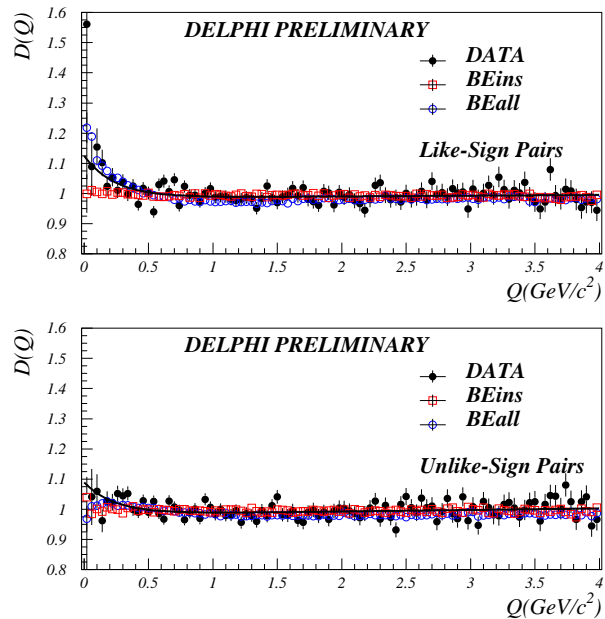}}   
\caption{The DELPHI $D(Q)$ plots (see text).}
\end{figure}

\vspace*{-1.cm}

\subsection{The L3 results} 

The L3 results using all available statistics are presented in~\cite{l3be}.
The measured value of $\Lambda$(L3) for like--sign pairs is consistent 
with zero, i.e. absence of inter--W correlations: 
$\Lambda$(L3)$ =0.008 \pm 0.018 (stat) \pm 0.012 (syst).$ 
The value for model with full BEC L3 quotes is
$\Lambda(full~BEC)$=0.098$\pm$0.008. 
L3 measured also the $\Delta\rho(Q)$ distributions. 

The value $\Lambda$(L3)
 has been obtained 
from the fits of the $D'(Q)$ plot in range of $Q$=0--1.4 GeV \label{beq}
by the equation 
$D'(Q) = N \left( 1 + \delta Q \right) \left( 1 + \Lambda e^{-k^2 Q^2} \right)$.
We noted that the two intriguing assumptions have been 
made in ~\cite{l3be}: (a) the parameter $k$ was fit parameter,
(b) the normalization $N$ was 
fixed to unity which does not allow the number of pairs to vary.
We performed the new test fits to L3 data plots in range of $Q$=0--1.4 GeV
(using the values and errors as presented in ~\cite{l3be}) to estimate 
the effects of (a) and (b).  We found the remarkable
differences in values of $\Lambda$, 
particulary, the statistical error in case of UNFIXED $N$ was significantly
higher than in case of fixed $N$=1, also significantly higher
than it is presented in published 
number of $\Lambda$(L3)$ =0.008 \pm 0.018 (stat) \pm 0.012 (syst).$  
 It is important particulary for comparison with model predictions.

\section{Summary}

The LEP data and theoretical predictions suggest that
the observed correlations between both, like-sign and 
unlike-sign particle pairs at small relative momenta are due to BEC.

The results by DELPHI~\cite{newnote} indicate clearly (2.8$\sigma$ effect
for like-sign and 2.0$\sigma$ effect for unlike-sign particle pairs) the
presence of BEC between particles from different Ws
in  $e^+e^-\rightarrow W^+W^-$ events, while the L3 
analysis in ~\cite{l3be} shows no effect. 
The DELPHI and L3 analysis use all available 
statistics and the event mixing method which, as generaly agreed, is 
the best method for the measuring of this effect.
We have found that there are uncertainties in main measured 
quantities in L3 analyses~\cite{l3be}. No WW BEC measurements exist 
yet for ALEPH and OPAL using all available statistics. 

Although there were many attempts to implement various approaches
of BEC effect in the models, LUBOEI model~\cite{PYTHIA} is the only 
BEC model the predictions of which were compared with the LEP correlation 
data in detail. The model describes the data reasonably well.
The latest version of the model, the LUBOEI BE$_{32}$ code~\cite{PYTHIA} 
with BEC included between particles from different Ws in WW fully 
hadronic channel yields about --35 MeV shift of W-mass due to BEC.

To quote Prof. Bo Andersson (Datong, 2001): if pions from different
Ws are produced close in phase space, they should show BEC, whether
or not coming from the same or different Ws; that is Quantum Mechanics.



\begin{thebibliography}{99}
\itemsep 0pt
\parsep 0pt
\bibitem{lafferty} G.~Lafferty, Z.\ Phys.\ {\bf C60} (1993) 659.
\bibitem{PYTHIA} T.~Sj\"ostrand et al., Comp.\ Phys.\ Comm.\ 135 (2001) 238.
L.\ L\"onnblad and T.\ Sj\"ostrand, Eur. Phys. J. {\bf C2} (1998) 165.
L.\ L\"onnblad and T.\ Sj\"ostrand, Phys.\ Lett.\ {\bf B351} (1995) 293.
\bibitem{newnote}  DELPHI Coll., DELPHI note 2002--032 CONF 566.
\bibitem{l3be} L3 Coll., L3 preprint 257.  Accepted by Physics Letters B.

\end{thebibliography}
\end{document}